\documentclass{emulateapj}
\usepackage{graphicx}
\usepackage{layout}
\usepackage[latin1]{inputenc}
\usepackage{amssymb}
\usepackage{natbib}

\slugcomment{To appear in ApJ}
\shorttitle{The peculiar optical spectrum of 4C+22.25}
\shortauthors{Decarli et al.}

\def\Msun{M$_\odot$}
\def\Mbh{$M_{\rm BH}$}

\def\Mhost{$M_{\rm host}$}

\def\Mgii{Mg\,{\sc ii}}

\def\Nev{[Ne\,{\sc v}]}
\def\Neiii{[Ne\,{\sc iii}]}
\def\Heii{He\,{\sc ii}}
\def\Oiii{[O\,{\sc iii}]}
\def\Oii{[O\,{\sc ii}]}
\def\Ha{H$\alpha$}
\def\Hb{H$\beta$}
\def\Hg{H$\gamma$}
\def\lsim{\mathrel{\rlap{\lower 3pt \hbox{$\sim$}} \raise 2.0pt \hbox{$<$}}}
\def\gsim{\mathrel{\rlap{\lower 3pt \hbox{$\sim$}} \raise 2.0pt \hbox{$>$}}}

\begin{document}

\title{
The peculiar optical spectrum of 4C+22.25: Imprint of a massive black hole 
binary?
}

\author{
R. Decarli\altaffilmark{1}, M. Dotti\altaffilmark{2}, C. Montuori\altaffilmark{3},
T. Liimets\altaffilmark{4,5}, A. Ederoclite\altaffilmark{6,7}
}
\altaffiltext{1}{Max-Planck Institut f\"{u}r Astronomie, K\"{o}nigstuhl 17, D-69117, Heidelberg, Germany. E-mail: {\sf decarli@mpia.de}}
\altaffiltext{2}{Max-Planck-Institut f\"{u}r Astrophysik, Karl-Schwarzschild-Str. 1, D-85748, Garching, Germany. E-mail: {\sf mdotti@mpa-garching.mpg.de}}
\altaffiltext{3}{Dipartimento di Fisica e Matematica, Universit\`a dell'Insubria,
via Valleggio 11, I-22100 Como, Italy}
\altaffiltext{4}{Nordic Optical Telescope, Apartado 474, E-38700 Santa Cruz de La Palma, 
Santa Cruz de Tenerife, Spain}
\altaffiltext{5}{Tartu Observatory, T\~oravere, 61602, Estonia.}
\altaffiltext{6}{Instituto de Astrofísica de Canarias, E-38200 La Laguna, Tenerife, Spain}
\altaffiltext{7}{Departamento de Astrofísica, Universidad de La Laguna, E-38205 La Laguna, Tenerife, Spain}


\begin{abstract}
We report the discovery of peculiar features in the optical spectrum of
4C+22.25, a flat spectrum radio quasar at $z=0.4183$ observed in the
SDSS and in a dedicated spectroscopic follow-up from the Nordic Optical 
Telescope. The \Hb{} and \Ha{} lines show broad profiles 
(FWHM$\sim$12,000 km s$^{-1}$), faint fluxes and extreme offsets 
($\Delta v=8,700\pm1,300$ km s$^{-1}$) with respect to the narrow 
emission lines. These features show no significant variation in a time lag 
of $\sim3.1$ yr (rest frame). We rule out possible interpretations based
on the superposition of two sources or on recoiling black holes, and we discuss
the virtues and limitations of a massive black hole binary scenario.
\end{abstract}

\section{Introduction}

4C+22.25 (RA: 10:00:21.8 Dec: +22:33:19  (J2000.0))
is a flat spectrum radio quasar at $z=0.4183$, discovered through
radio observations by \citet{merkelijn68}. A first optical spectrum 
was collected by \citet{schmidt74} who observed a flat continuum with no 
significant emission line, suggesting that the source is a BL Lac object. 
\citet{haddad91} re-observed the 4000--6000 \AA{} range and detected a set 
of bright, marginally resolved narrow lines (\Nev{}$_{3346,3426}$; 
\Neiii{}$_{3869,3968}$; \Oii{}$_{3727}$; \Hg{}, \Oiii{}$_{4363}$) at 
$z=0.419$. The intensity ratios of these lines suggested that the source 
hosts a Seyfert-like Narrow Line (NL) region. \citet{nilsson03} collected 
ground-based high-resolution images of 4C+22.25 as a part of a study of 
blazar host galaxies, and showed that the host galaxy is well 
resolved. Its light profile follows a de Vaucouleurs law with scale radius 
$R_e=3.3\pm0.2''$ ($18\pm1$ kpc) and apparent magnitude $m_R=18.63\pm0.05$. 
Including a $k$-correction ($0.7$ mag assuming a typical Elliptical galaxy 
spectrum at $z=0.419$), the inferred luminosity is $M_R=-23.9$.

A companion galaxy is located $\sim6''$ North-West of 4C+22.25 ($\sim30$ 
kpc at the redshift of the quasar). \citet{haddad91} reported the detection 
of the CaII (H) and G-band features in its spectrum, yielding $z=0.416$,
and suggested that a gravitational interaction with the quasar host galaxy
may be occurring. 
SDSS photometry reveals also the presence of $12$ galaxies within
a projected separation of 400 kpc and photometric redshift consistent with
the one of 4C+22.25, suggesting that the source may be located in a 
relatively rich galactic environment.

In this Letter we report the discovery of peculiar broad lines in the 
optical spectrum of 4C+22.25 that is publicly available from the Sloan
Digital Sky Survey \citep[SDSS;][]{york00} database. Very broad and 
rather faint \Mgii{}, \Hb{} and \Ha{} lines are observed, all showing a 
velocity blueshift of $\approx8,700$ km s$^{-1}$ with respect to the NLs.

Similar velocity offsets have been already observed in a handful of 
SDSS quasars: \citet{komossa08} reported a $\sim2,650$ km s$^{-1}$ shift
between the main narrow line system and a second set of narrow and broad 
lines in SDSS J092712.65+294344.0 (hereafter, J0927), which has been 
interpreted as the signature of a recoiling black hole \citep{komossa08}, 
of a massive black hole binary \citep[BHB;][]{bogdanovic09,dotti09} or the 
superposition of two objects \citep{heckman09}. Similarly, \citet{shields09}
found a $\sim3,500$ km s$^{-1}$ shift between narrow and broad lines in the 
spectrum of SDSS J105041.35+345631.3 (hereafter, J1050). Finally, 
\citet{boroson09} revealed the presence of a peculiar profile
in the broad lines of SDSS J153636.22+044127.0, which could be
due to a BHB \citep{boroson09,lauer09}, a superposition of quasars
\citep{wrobel09,decarli09}, or an extreme double-peaked emitter 
\citep{tang09,chornock10}. However, we show here that most of these 
interpretations are unsuitable for 4C+22.25. 

The structure of the Letter is the following: 
in Section \ref{sec_spectra} we analyse the SDSS spectrum and present
new observations collected at the Nordic Optical Telescope
(NOT). In Section \ref{sec_discussion} we discuss possible models
to interpret the peculiar features of this source. Conclusions
are summarized in Section \ref{sec_conclusions}. Throughout the Letter
we will assume a standard cosmology with $H_0=70$ km s$^{-1}$ Mpc$^{-1}$, 
$\Omega_{\rm m}=0.3$ and $\Omega_{\Lambda}=0.7$.

\section{The spectroscopic observations}\label{sec_spectra}

\subsection{SDSS spectrum}

The SDSS spectrum of 4C+22.25 was collected on January, 2$^{\rm nd}$, 
2006, and was published in the SDSS Sixth Data Release \citep{sdssdr6}.
SDSS spectra have $\lambda/\Delta \lambda\sim 2000$ and cover the 
$3800$--$9000$ \AA{} range. Uncertainties on wavelength calibration amount to 
$0.05$ \AA, while flux calibration formal errors account to 5\%. 
The signal-to-noise ratio per pixel at 6400 \AA{} is 21.

\begin{figure}
\begin{center}
\includegraphics[width=0.5\textwidth]{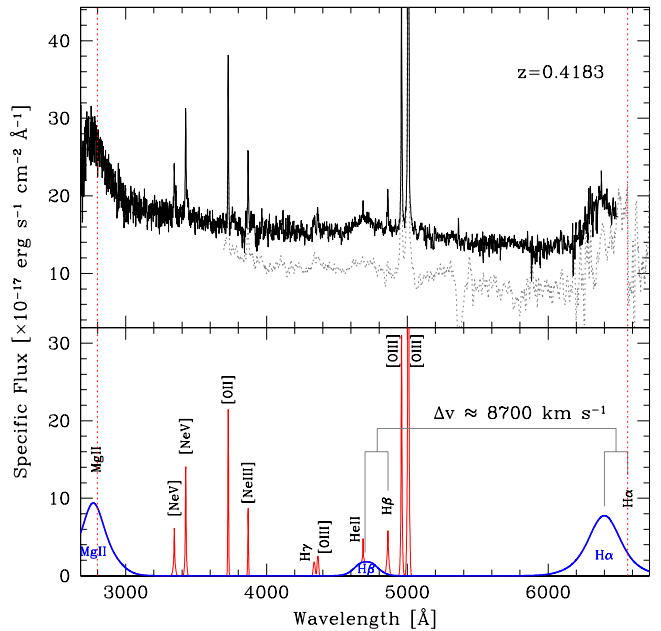}\\
\caption{{\it Top panel:} SDSS (solid line) and NOT (dotted line) spectra of 
4C+22.25, shifted to the rest frame assuming $z=0.4183$. The NOT spectrum
is shifted downwards for the sake of clarity. No significant
difference in the two spectra is reported. {\it Bottom panel:} 
the fitted components. Thick lines mark the broad lines, thin lines mark the 
narrow features. The two vertical dotted lines show the rest-frame wavelengths 
of \Mgii{} and \Ha{}. The \Mgii{} line model is poorly constrained since the 
line lays at the edge of the SDSS coverage. The velocity shift between the broad
and the narrow line systems is clearly apparent.}\label{fig_spc}
\end{center}
\end{figure}

\begin{figure}
\begin{center}
\includegraphics[width=0.5\textwidth]{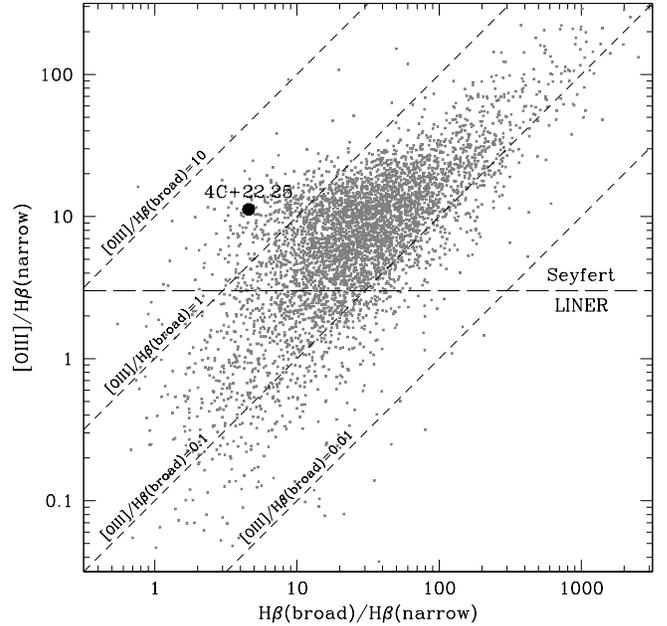}\\
\caption{The \Oiii{}/\Hb{}(narrow) flux ratio plotted as a function of the
ratio between broad and narrow \Hb{} components. The horizontal line
marks the separation between Seyfert and LINER ionization conditions 
\citep[see][]{heckman80}. Diagonal lines show the loci of various 
\Oiii{}/\Hb{}(broad) flux ratios. The big dot marks the position of 
4C+22.25. For reference, we plot with small grey dots the flux ratios of
the 18,101 quasars from the compilation by \citet{shen10} for which the 
three lines were detected. 4C+22.25 shows strong \Oiii{} and very faint
broad \Hb{} as compared to the average quasar population.}\label{fig_fluxes}
\end{center}
\end{figure}

Figure \ref{fig_spc} shows the SDSS spectrum (top panel) and the identification
of main emission lines (bottom panel). Emission lines
were fitted with a double-gaussian profile, following \citet{decarli08}.
Relevant information are provided in Table \ref{tab_lines}.
Typical uncertainties in the line FWHM are around 10\%. NL peak wavelengths
have uncertainties of 10 to few hundred km s$^{-1}$, depending on the line
flux (see Table \ref{tab_lines}). The broad line peak wavelengths are poorly 
constrained: for \Hb{} and \Ha{}, we estimate recessional velocity 
uncertainties of $1,900$ and $1,700$ km s$^{-1}$ respectively.
The narrow lines are marginally resolved. Their mean redshift is 
$\langle z \rangle = 0.4183$.
The \Oiii{}/\Oii{}, \Oiii{}/\Hb{} and \Nev{}/\Neiii{} flux ratios 
confirm the presence of Seyfert-like ionization conditions in the NL region
\citep[see Figure \ref{fig_fluxes} and][]{heckman80,haddad91}.

For the first time, we report the detection of broad lines in 4C+22.25. The
\Ha{}, \Hb{} and \Mgii{} lines are clearly visible, while broad components
of other Balmer lines and the iron multiplets are too faint to be detected. 
Both \Ha{} and \Hb{} are very broad \citep[FWHM$\sim$12,000 km s$^{-1}$, i.e., larger
than 96\% of the quasars in the huge, SDSS-based dataset by][]{shen10} and faint with
respect to, e.g., the narrow \Oiii{} lines \citep[see Figure \ref{fig_fluxes}; 
only 1.1\% of the quasars in][has larger \Oiii{}/\Hb{}(broad) values]{shen10}. The most striking property of 
these lines is that they show enourmous blueshifts ($8,700\pm1,300$ km s$^{-1}$) 
with respect to the narrow-line system. Similar properties (in terms of fluxes, line width 
and shift) are reported also for the \Mgii{} line, but since the peak is close
to the range covered by the SDSS spectrum, the line characterization is 
not feasible with the available data. We use the line width and luminosity 
of broad \Hb{} to compute the mass of the active black hole, following 
\citet{vestergaard06}: \Mbh{}=$1\times10^{9}$ \Msun{}. Assuming the bolometric
correction factor by \citet{richards06} for the continuum luminosity at 5100 
\AA, this yields $L/L_{\rm Edd}=0.035$.

\begin{table}
\caption{{\rm Summary of the main emission lines as observed in the 
SDSS and NOT spectra of 4C+22.25. (1) Line identification. When both 
broad and narrow components are available, they are marked with `b' and `n'
respectively. (2) Observed peak wavelength. (3) Redshift corresponding to 
the observed peak wavelength. (4) Velocity difference with respect to the 
mean redshift of the narrow line system, $\langle z \rangle = 0.4183$. 
Negative values correspond to blueshifts. (5) Full Width at Half Maximum 
of the fitted lines. Note that no correction for spectral resolution is
applied here. (6) Line luminosity. }} \label{tab_lines}   
\begin{center}
\begin{tabular}{cccccc}
   \hline             
    Line       & $\lambda_{\rm obs}$ &  $z$  &  $\Delta v$ &  FWHM & log $L_{\rm line}$ \\
 	     	  &  [\AA]  &	     &  [km s$^{-1}$] & [km s$^{-1}$]&   [erg s$^{-1}$]   \\
     (1)     	  &   (2)   &  (3)   &   (4)         &   (5) &    (6)      \\
   \hline 
   \multicolumn{6}{l}{\it SDSS spectrum}                                   \\
    \Nev{}     	  & 4745.8  & 0.4180 &  $-60\pm720$  &   580 &  41.60      \\
    \Nev{}     	  & 4860.2  & 0.4183 &  $-10\pm120$  &   560 &  41.87      \\
    \Oii{}     	  & 5288.2  & 0.4183 &  $+10\pm50$   &   490 &  41.96      \\
  \Neiii{}     	  & 5488.9  & 0.4184 &  $+20\pm160$  &   410 &  41.53      \\
  \Hg{}(n)        & 6155.4  & 0.4178 & $-120\pm600$  &  1100 &  41.20      \\
   \Oiii{}     	  & 6192.3  & 0.4188 & $+110\pm500$  &   910 &  41.36      \\
   \Heii{}    	  & 6647.4  & 0.4183 &  $-10\pm300$  &   280 &  41.08      \\
  \Hb{}(b)        & 6679.6  & 0.3736 & $-9700\pm1900$& 12000 &  42.34	   \\
  \Hb{}(n)  	  & 6897.4  & 0.4184 &  $+30\pm130$  &   570 &  41.68      \\
   \Oiii{}     	  & 7035.2  & 0.4183 &    $0\pm30$   &   430 &  42.25      \\
   \Oiii{}     	  & 7103.0  & 0.4183 &   $-8\pm11$   &   410 &  42.73      \\
  \Ha{}(b)        & 9078.8  & 0.3830 & $-8000\pm1700$& 12700 &  43.18      \\
  \hline
   \multicolumn{6}{l}{\it NOT spectrum}                                    \\
  \Hb{}(b)        & 6681.6  & 0.3740 & $-9400\pm1600$& 13000 &  42.32	   \\
  \Oiii{}     	  & 7104.2  & 0.4185 &  $+50\pm30$   &   770 &  42.74      \\
  \hline
\end{tabular}
   \end{center}
\end{table}

Few absorption features are also tentatively reported, namely the \Mgii{} 
doublet, the Ca (H) and (K) and the NaD line, at a redshift consistent with 
the narrow emission lines.

\subsection{NOT spectrum}

We re-observed 4C+22.25 using the Andalucia Faint Object Spectrograph and
Camera (ALFOSC) mounted on the 2.56m Nordic Optical Telescope (NOT) on
June, 2$^{\rm nd}$, 2010, i.e., 1,612 days after the acquisition of the SDSS
spectrum ($1,137$ days in the rest frame of the source). Long-slit 
spectroscopy configuration was adopted. Grism \# 5 yields a spectral
resolution $\lambda/\Delta \lambda\approx410$ ($1.0''$ slit) in the
spectral range 5,500--10,000 \AA{}. The total integration time (45 min)
was split into 3 exposures to allow an easy cleaning of cosmic rays.
Standard IRAF tools were used to reduce data. Wavelength calibration was 
performed using Th-Ar arc spectra, and cross-checked using the sky emission 
lines in the science spectra. Wavelength residual rms is $1.3$ \AA{}. Flux 
calibration was achieved observing a spectrophotometric standard star. 
Galactic extinction was accounted for according to \citet{schlegel98}, 
assuming $R_V = 3.1$. The final spectrum is shown in Figure \ref{fig_spc}, 
top panel. Its signal-to-noise per pixel at 6400 \AA{} is 23. The NOT 
spectrum is in excellent agreement with the SDSS observation both in terms of 
fluxes and peak wavelengths of the observed features (see Table 
\ref{tab_lines}). Since the SDSS spectrum has a better global quality, we 
will refer to velocities and fluxes derived from the SDSS data in the 
following analysis.

\section{Discussion}\label{sec_discussion}

\subsection{What 4C+22.25 cannot be}

A simple explanation of the two redshift systems observed in 4C+22.25 
would be that the broad and narrow emission lines belong to two different 
objects, superimposed along the line of sight. This scenario is 
disfavoured by the lack of narrow emission lines at the redshift of the
broad line system \citep[see also][]{boroson09}. We estimate that
a narrow line as faint as $3.6 \times 10^{40}$ erg s$^{-1}$ would be detected
at 1-sigma with respect to the noise of the SDSS spectrum. This limit
corresponds to $0.017 \times$ the flux of the broad component of \Hb{}. From
Figure \ref{fig_fluxes}, it is apparent that the number of quasars with 
\Oiii{}/\Hb{}(broad)$<0.017$ is negligible.

Moreover, in order to get both the sources within the fiber aperture of the 
SDSS, their separation should be $<1.5''$, yielding a solid angle 
$<5.5\times10^{-7}$ deg$^{2}$. The number density of Active Galactic Nuclei 
(AGN) at $0.35<z<0.45$ (i.e., in a velocity space three times as large as 
the velocity offset observed in 4C+22.25) is $\sim0.37$ deg$^{-2}$ 
\citep{schneider10}. Hence, the probability of having a random superposition
is $\sim 2 \times 10^{-7}$, i.e., completely negligible if compared to the 
number of SDSS AGN in this redshift bin ($\sim3,300$). The probability 
of alignment of two AGN substantially increases if they belong to a common 
physical structure, e.g., a cluster of galaxy. This scenario was proposed 
by \citet{heckman09} to interpret the two redshift systems observed in 
J0927, but subsequent observations revealed that no significant cluster is 
present \citep{drd09}.
The `superposition in a cluster' argument cannot be applied to 4C+22.25, 
as the velocity difference between the two line systems is too high to be 
attributed to the potential well of a single physical structure
\citep[see the statistical analysis by][]{dotti10a}. 

Another scenario suggested to explain the velocity shifts between narrow
and broad lines observed in J0927 and J1050 is that the black hole in
these quasars is recoiling, as a result of the coalescence of a BHB 
\citep{komossa08,shields09}. The maximum recoil achievable during BH 
coalescence is $\lsim 4,000$ km s$^{-1}$
\citep{backer07,herrmann07,campanelli07,schnittman07,lousto09,vanmeter10}\footnote{Note
that hydrodynamical and/or purely relativistic effects can strongly
suppress the kick magnitude \citep{schnittman04,bogdanovic07,dotti10b,volonteri10,kesden10}.}.  
As a consequence, the recoiling scenario is ruled out for 4C+22.25.

\subsection{What 4C+22.25 might be}

A possible alternative is that 4C+22.25 hosts a binary black hole. In this 
picture the primary, more massive BH resides at the center of a circumbinary 
gaseous disk, located in the nuclear region of the host galaxy, while
a secondary black hole orbits around it. Because of its motion, the 
secondary black hole simultaneously accretes and prevents the primary one 
from accreting. The velocity shift between narrow and broad lines is then
due to the Keplerian velocity of the secondary black hole 
with respect to the barycenter of the binary 
\citep[for more details, see e.g.][]{bogdanovic09,dotti09}. 
Assuming circular orbits, the orbital period $t$ would be:
\begin{equation}\label{eq_period}
t = 2 \pi \frac{G M_2 (\sin \vartheta \cos \phi)^3}{q(1+q)^2 (\Delta v)^3}
\end{equation}
where $M_{1,2}$ is the mass of the primary and secondary black holes, $q=M_2/M_1$, 
$\vartheta$ is the inclination angle of the rotational axis with respect to the 
line of sight, and $\phi$ is the orbital phase (defined so that 
$\phi=0$ at the orbital node maximizing the blueshift of the broad lines). Similarly, 
the separation $a$ between the two black holes would be:
\begin{equation}\label{eq_separation}
a= \frac{G M_2 (\sin \vartheta  \cos \phi)^2}{q (1+q) (\Delta v)^2}
\end{equation}
In order to charachterize the properties of the BHB, we therefore need
an estimate of $M_1$ and $M_2$, which are unknown. 
Following \citet{decarli10}, we use the host luminosity to infer the
expected mass of $M_1$, assuming \Mbh{}/\Mhost{}=$0.0015$ as observed 
in the Local Universe \citep[e.g.,][]{marconi03}. For an old host galaxy 
stellar population, we infer $M_1=2 \times 10^9$ \Msun{}.
Assuming $M_2=1\times10^9$ \Msun{}, as derived in Section \ref{sec_spectra},
we obtain separations of 0.04--0.08 pc and orbital periods of 15--35 years 
for $\vartheta=45^{\circ}-90^{\circ}$ and $\phi=0$. On the other hand,
the velocity shift observed in the SDSS and the NOT spectra is unchanged
within the uncertainties ($\sim 2,000$ km s$^{-1}$). This implies that 
the period should be $\gsim30$ yr. 

Small but not negligible eccentricities are expected in very massive 
BHBs, driven by three body interacitons with stars. For $q\sim1$ (as in 
the present case), the maximum expected eccentricity is 0.1--0.3, depending
on the mass of the binary and the steepness of the radial distribution
of stars in the host galaxy \citep[see][]{sesana10}. Such small 
eccentricities do not change significantly our estimates\footnote{Higher 
eccentricities, besides being disfavoured by models, are also ruled out by 
the absence of a velocity shift in the two observations. In a very 
eccentric orbit, the secondary black hole spends most of its time close to 
the apocenter, where the velocity has to be larger than (or
equal to) the velocity observed in the spectrum ($\Delta v$). This implies
that the period of the eccentric binary would be much shorter, hence
incompatible with the observational constraints.}

We point out that, at these tiny separations, the broad line region is 
expected to be perturbed. This would explain the faintness of the broad 
lines with respect to the narrow lines (see Figure \ref{fig_fluxes}). 
In this case, the \citet{vestergaard06} recipe used to estimate $M_2$ may 
not be valid. We therefore adopt a different role-of-thumb approach
to estimate $M_2$, namely assuming that the quasar is accreting at 10\% of
its Eddington luminosity. In this case, $M_2\approx3\times10^8$ \Msun{}, 
$a\approx0.05-0.1$ pc and $t\approx20-60$ yr. We conclude that the BHB 
scenario is a viable one for 4C+22.25.

\section{Conclusions}\label{sec_conclusions}

We present the discovery of extremely peculiar features in the optical
spectrum of the flat spectrum radio quasar 4C+22.25. The narrow lines are 
very bright and reveal the presence of a Seyfert-like nucleus. Its broad 
lines are faint and flat (FWHM$\sim$12,000 km s$^{-1}$), and blueshifted 
with respect to the NL of $8,700\pm1,300$ km s$^{-1}$. This velocity offset 
between broad and NLs is so high that scenarios involving a superposition 
in a cluster or a recoiling black hole are ruled out at high confidence. 
The probability of a chance superposition of two AGN on cosmological scales 
is so small that it disfavored, especially if coupled with the 
non-detection of any narrow emission line at the redshift of the broad line
system. The massive black hole binary scenario holds for 4C+22.25, but the 
observation of the target in two different epochs separated by $3.1$ yr
(rest frame) allowed us to set strong constraints on the possible 
orbital configurations. New observations with a longer time lag will help 
clarifying if the binary model is correct or not. Moreover, observations
at higher frequencies, e.g., in the X-rays, would help in constraining
the mass and Eddington rate of the accreting black hole.

Whether 4C+22.25 is a lone object, or just an extreme case of a new subclass
of AGN, including J0927 and J1050, is not clear, and demands further 
investigation both from a theoretical and an observational point of view.

\section*{Acknowledgments}

We thank Fabian Walter, Alessia Gualandris and the anonymous referee 
for fruitful comments and discussions and Yue Shen for 
kindly making his catalogue available before publication.
Based on observations made with the Nordic Optical Telescope, operated
on the island of La Palma jointly by Denmark, Finland, Iceland,
Norway, and Sweden, in the Spanish Observatorio del Roque de los
Muchachos of the Instituto de Astrofisica de Canarias. 
The data presented here have been taken using ALFOSC, which is owned by the Instituto 
de Astrofisica de Andalucia (IAA) and operated at the Nordic Optical Telescope under 
agreement between IAA and the NBIfAFG of the Astronomical Observatory of Copenhagen.
This research has made use of the NASA/IPAC Extragalactic Database (NED)
which is operated by the Jet Propulsion Laboratory, California
Institute of Technology, under contract with the National Aeronautics
and Space Administration. 

Facilities:
\facility{NOT(ALFOSC)}
\facility{SDSS}

\label{lastpage}

\end{document}